\documentclass[11pt]{article}%
\usepackage{amsfonts}
\usepackage{amsmath}
\usepackage{amssymb}
\usepackage{graphicx}%
\providecommand{\U}[1]{\protect\rule{.1in}{.1in}}
\newtheorem{theorem}{Theorem}

\begin{document}

\title{WEAK\ VALUES\ OF\ A\ QUANTUM\ OBSERVABLE AND\ THE\ CROSS-WIGNER DISTRIBUTION}
\author{Maurice A. de Gosson\\University of Vienna\\Faculty of Mathematics, NuHAG\\A-1090 Vienna
\and Serge M. de Gosson\\Swedish Social Insurance Agency\\Department for Analysis and Forecasts\\103 51 Stockholm}
\maketitle

\begin{abstract}
We study the weak values of a quantum observable from the point of view of the
Wigner formalism. The main actor is here the cross-Wigner transform of two
functions, which is in disguise the cross-ambiguity function familiar from
radar theory and time-frequency analysis. It allows us to express weak values
using a complex probability distribution. We suggest that our approach seems
to confirm that the weak value of an observable is, as conjectured by several
authors, due to the interference of two wavefunctions, one coming from the
past, and the other from the future.

\end{abstract}

\noindent\textbf{PACS:} 03.65.-w, 02.30.Nw, 02.50.Cw, 03.65.Ta

\noindent\textbf{Keywords:} weak values, weak measurements, Wigner
distribution, interference

\section{Introduction}

We study in the present Letter the notion of weak measurement introduced by
Aharonov and Albert \cite{aav}, Bergmann, and Lebowitz in
\cite{abl,aav,avphysa,av,ahbo,apt} from the point of view of the Wigner phase
space formalism. This will allow us to discuss the claim made by these authors
that the weak value can be seen as the interference of two wavepackets, one
going forward in time and the other backwards in time.

Let us briefly recall the difference between an ideal (also called strong, or
von Neumann) measurement, and a weak measurement. Let $\widehat{A}$ be a
(quantum) observable, realized as an essentially self-adjoint operator; we
assume for simplicity that $\widehat{A}$ has a eigenvalues $a_{1},a_{2},...$
with corresponding orthogonal eigenfunctions $\psi_{1},\psi_{2},...$ . In an
ideal measurement the expectation value of $\widehat{A}$ in a pre-selected
state $\psi$ is
\begin{equation}
\langle\widehat{A}\rangle^{\psi}=\frac{\langle\psi|\widehat{A}|\psi\rangle
}{\langle\psi|\psi\rangle}; \label{expect}%
\end{equation}
if the sequence of eigenvalues lies in some interval $[a_{\min},a_{\max}]$
then we will have $a_{\min}\leq\langle\widehat{A}\rangle^{\psi}\leq a_{\max}$.
In fact, if one performs the ideal measurement the outcome will always be one
of the eigenvalues $\lambda_{j}$, and the probability of this outcome is
$|\lambda_{j}|^{2}/||\psi_{j}||^{2}$ where $\lambda_{j}$ is the coefficient of
$\psi_{j}$ in the Fourier expansion $\psi=\sum_{j}\lambda_{j}\psi_{j}$.
Moreover the system will be left in the state $\psi_{j}$ after the ideal
measurement yielding the value $a_{j}$. The situation is very different for
weak measurements. As is explained in Ritchie et al. \cite{risora} (also see
Berry and Shukla \cite{berry}, Steinberg \cite{stein}), in a weak measurement
the eigenvalues are not fully resolved and the system is left in a
superposition of the unresolved states. If an appropriate post-selection is
made, this superposition can interfere to produce a measurement result which
can be significantly outside the range of the eigenvalues of the observable
$\widehat{A}$. The post-selection can then be accomplished by making an ideal
measurement of some other observable $\widehat{B}$ and selecting one
particular outcome. Thus, the post-selected state $|\phi\rangle$ is an
eigenstate of $\widehat{B}$ which can be expressed as a linear combination of
the eigenstates of $\widehat{A}$ (we note that, conversely, an ideal
measurement can be expressed as a convex sum of weak values: see Hosoya and
Shikano \cite{hoyu}). If $\langle\phi|\psi\rangle\neq0$ (and if $\phi$, $\psi$
are square integrable) the \emph{weak value} of $\widehat{A}$ is then the
complex number%
\begin{equation}
\langle\widehat{A}\rangle_{\mathrm{weak}}^{\phi,\psi}=\frac{\langle
\phi|\widehat{A}|\psi\rangle}{\langle\phi|\psi\rangle}. \label{abl}%
\end{equation}

We will show that this weak value can be expressed in terms of the
cross-Wigner transform
\begin{equation}
W(\phi,\psi)(x,p)=\left(  \tfrac{1}{2\pi\hbar}\right)  ^{N}\int_{\mathbb{R}%
^{N}}e^{-\frac{i}{\hbar}py}\phi^{\ast}(x+\tfrac{1}{2}y)\psi(x-\tfrac{1}%
{2}y)dy\label{crosswigner}%
\end{equation}
of the pair $(\phi,\psi)$ whose physical interpretation is that of an
interference term in the Wigner distribution of the sum $\phi+\psi$; we
mention that the importance of these interference terms have been emphasized
and studied by Zurek \cite{zurek} in the context of the sub-Planckian
structures in phase space. The cross-Wigner transform is a very important
object being intensively studied in the harmonic analysis literature and in
time-frequency analysis; see e.g. Cohen \cite{Cohen}, Folland \cite{Folland}
Gr\"{o}chenig \cite{Gro}, Hlawatsch and Flandrin \cite{hlafla}. Notice that
$W(\phi,\psi)(x,p)$ reduces to the familiar Wigner distribution (Hillery et
al. \cite{hiscuwi}, Littlejohn \cite{Littlejohn} when $\phi=\psi$. 

We will not address here the ontological debates arising around the problem of
\textquotedblleft Elements of Reality\textquotedblright\ (see Cohen and Hiley
\cite{cohi1,cohi2}); these questions are difficult and have led to profound
philosophical controversies.\bigskip

\noindent\textbf{Notation.} We will work with systems having $N$ degrees of
freedom. Position (resp. momentum) variables are denoted $x=(x_{1},...,x_{N})$
(resp. $p=(p_{1},...,p_{N})$); they are vectors in $\mathbb{R}^{N}$. The
corresponding phase space variable is $z=(x,p)$; it is a vector in phase space
$\mathbb{R}^{2N}$. We will endow the phase space with the standard symplectic
form $\sigma(z,z^{\prime})=px^{\prime}-p^{\prime}x$. When integrating we will
use, where appropriate, the volume elements $dx=dx_{1}\cdot\cdot\cdot dx_{N}$,
$dp=dp_{1}\cdot\cdot\cdot dp_{N}$, $dz=dpdx$. The unitary $\hbar$-Fourier
transform of a function $\psi$ in $L^{2}(\mathbb{R}^{N})$ is defined by%
\[
F\psi(p)=\left(  \tfrac{1}{2\pi\hbar}\right)  ^{N/2}\int_{\mathbb{R}^{N}%
}e^{-\frac{i}{\hbar}py}\psi(y)dy.
\]

\section{The Main Result}

\subsection{A complex probability distribution}

The cross-Wigner transform (\ref{crosswigner}) satisfies the \textquotedblleft
marginal properties\textquotedblright%
\begin{equation}
\int_{\mathbb{R}^{N}}W(\phi,\psi)(z)dp=\phi(x)^{\ast}\psi(x) \label{psifi}%
\end{equation}
and%
\begin{equation}
\int_{\mathbb{R}^{N}}W(\phi,\psi)(z)dx=F\phi(p)^{\ast}F\psi(p). \label{psifif}%
\end{equation}
It follows from the equality (\ref{psifi}) that%
\begin{equation}
\int_{\mathbb{R}^{N}}W(\phi,\psi)(z)dz=\langle\phi|\psi\rangle. \label{fipsi}%
\end{equation}

For $\langle\phi|\psi\rangle\neq0$ we define%
\begin{equation}
\rho_{\phi,\psi}(z)=\frac{W(\phi,\psi)(z)}{\langle\phi|\psi\rangle}.
\label{ropsifi}%
\end{equation}
Note the conjugation relation $\rho_{\phi,\psi}(z)^{\ast}=\rho_{\psi,\phi}%
(z)$; also $\rho_{\lambda\phi,\lambda\psi}(z)=\rho_{\phi,\psi}(z)$ for every
complex $\lambda\neq0$ hence the function $\rho_{\phi,\psi}$ only depends on
the states $|\psi\rangle$ and $|\phi\rangle$. In view of Eqn. (\ref{fipsi}) we
have%
\begin{equation}
\int_{\mathbb{R}^{2N}}\rho_{\phi,\psi}(z)dz=1 \label{norm}%
\end{equation}
hence $\rho_{\phi,\psi}$ can be viewed as a complex probability distribution
with respect to which the average of the classical observable $A$ is
calculated; also, Eqn. (\ref{norm}) implies that%
\begin{equation}
\int_{\mathbb{R}^{2N}}\operatorname{Re}\rho_{\phi,\psi}(z)dz=1\text{ \ ,
\ }\int_{\mathbb{R}^{2N}}\operatorname{Im}\rho_{\phi,\psi}(z)dz=0
\label{normbis}%
\end{equation}
so that $\operatorname{Re}\rho_{\phi,\psi}$ can be viewed as a
quasi-distribution, in the same way as the usual Wigner transform. When
$\psi=\phi$ then $\operatorname{Im}\rho_{\psi,\psi}=0$ and $\operatorname{Re}%
\rho_{\psi,\psi}=W\psi$. Observe that it immediately follows from Eqns.
(\ref{ropsifi}) and (\ref{psifi}), (\ref{psifif}) that the marginals
distributions of $\rho_{\phi,\psi}$ are given by%
\begin{equation}
\int_{\mathbb{R}^{N}}\rho_{\phi,\psi}(z)dp=\frac{\phi^{\ast}(x)\psi
(x)}{\langle\phi|\psi\rangle}\text{ \ , \ }\int_{\mathbb{R}^{N}}\rho
_{\phi,\psi}(z)dx=\frac{[F\phi(p)]^{\ast}F\psi(p)}{\langle\phi|\psi\rangle};
\label{marginals}%
\end{equation}
note that anyone of these equalities allows by integrating in the conjugate
variable to recover the normalization condition (\ref{norm}).

We point out that the consideration of complex probability densities has
\textit{per se} nothing unusual; such complex probabilities have been used in
the context of stochastic processes (see Zak \cite{zak}), signal theory
(multipath fading channels, see Chayawan \cite{cha}) and they also appear in
the study of non-Hermitian quantum mechanics (see Barkay and Moiseyev
\cite{bar}).

We claim that:

\begin{theorem}
Let $A$ be a classical observable and $\widehat{A}$ its Weyl quantization; we
have%
\begin{equation}
\langle\widehat{A}\rangle_{\mathrm{weak}}^{\phi,\psi}=\int_{\mathbb{R}^{2N}%
}A(z)\rho_{\phi,\psi}(z)dz. \label{formula3}%
\end{equation}

\end{theorem}

The reader familiar with the Weyl--Wigner--Moyal formalism (de Gosson
\cite{Birk,Birkbis}, Littlejohn \cite{Littlejohn}) will have noticed that when
$\phi=\psi$ formula (\ref{formula3}) reduces to the well-known relation%
\[
\langle\psi|\widehat{A}|\psi\rangle=\int_{\mathbb{R}^{2N}}A(z)W\psi(z)dz
\]
yielding the usual expectation value $\langle\widehat{A}\rangle^{\psi}%
=\langle\psi|\widehat{A}|\psi\rangle/\langle\psi|\psi\rangle$. We will study
the relative importance of these values when $\phi$ and $\psi$ are coherent
states in Subsection \ref{subco}

\subsection{Proof of Theorem 1}

To prove formula (\ref{formula3}) it is sufficient, in view of definition
(\ref{ropsifi}) of $\rho_{\phi,\psi}(z)$, to show that
\begin{equation}
\langle\phi|\widehat{A}|\psi\rangle=\int_{\mathbb{R}^{2N}}W(\phi
,\psi)(z)A(z)dz. \label{formula2}%
\end{equation}
To prove the latter we could perform a direct calculation staring from the
right-hand side, inserting the expression (\ref{crosswigner}) of $W(\phi
,\psi)(z)$ and making various changes of variables. We prefer to give a more
elegant proof which has some conceptual advantages. The first step consists in
observing that the cross-Wigner transform can be expressed in terms of the
Grossmann--Royer \cite{Gros,Roy} operator
\begin{equation}
\widehat{T}_{\text{GR}}(z_{0})\phi(x)=e^{\frac{2i}{\hbar}p_{0}(x-x_{0})}%
\phi(2x_{0}-x) \label{GR}%
\end{equation}
(also see de Gosson \cite{Birkbis}, Chapter 9). A simple calculation shows
that we have
\begin{equation}
W(\phi,\psi)(z)=\left(  \tfrac{1}{\pi\hbar}\right)  ^{N}\langle\widehat{T}%
_{\text{GR}}(z)\phi|\psi\rangle\label{wigroyer}%
\end{equation}
and that the Weyl quantization $\widehat{A}$ of the observable is given by
\begin{equation}
\widehat{A}\psi(x)=\left(  \tfrac{1}{\pi\hbar}\right)  ^{N}\int_{\mathbb{R}%
^{2N}}A(z_{0})\widehat{T}_{\text{GR}}(z_{0})\psi(x)dz_{0}. \label{forweylgr}%
\end{equation}
Using the latter we have
\begin{equation}
\langle\phi|\widehat{A}|\psi\rangle=\left(  \tfrac{1}{\pi\hbar}\right)
^{N}\int_{\mathbb{R}^{2N}}A(z_{0})\langle\phi|\widehat{T}_{\text{GR}}%
(z_{0})\psi\rangle dz_{0}; \label{fiap}%
\end{equation}
we next observe that $\widehat{T}_{\text{GR}}(z_{0})$ is both unitary and
involutive (i.e. $\widehat{T}_{\text{GR}}(z_{0})=\widehat{T}_{\text{GR}}%
(z_{0})^{-1}$) and hence
\begin{equation}
\langle\phi|\widehat{T}_{\text{GR}}(z_{0})\psi\rangle=\langle\widehat{T}%
_{\text{GR}}(z_{0})^{-1}\phi|\psi\rangle=\langle\widehat{T}_{\text{GR}}%
(z_{0})\phi|\psi\rangle\label{figri}%
\end{equation}
so that (\ref{fiap}) can be rewritten
\begin{align*}
\langle\phi|\widehat{A}|\psi\rangle &  =\left(  \tfrac{1}{\pi\hbar}\right)
^{N}\int_{\mathbb{R}^{2N}}A(z_{0})\langle\widehat{T}_{\text{GR}}(z_{0}%
)\phi|\psi\rangle dz_{0}\\
&  =\int_{\mathbb{R}^{2N}}A(z_{0})W(\phi,\psi)(z)dz_{0}%
\end{align*}
which was to be proven.

\subsection{The case of coherent states\label{subco}}

Suppose that both wavefunctions are normalized coherent states concentrated
near $z_{0}=(x_{0},p_{0})$ and $-z_{0}$ at time $t_{\mathrm{in}}$, that is we
choose $\theta$ and $\psi=\psi_{z_{0}}$ where%
\begin{equation}
\theta(x)=\left(  \tfrac{1}{\pi\hbar}\right)  ^{N/4}\widehat{T}(z_{0}%
)e^{-\frac{1}{\hbar}|x|^{2}}\text{ \ , \ }\psi(x)=\left(  \tfrac{1}{\pi\hbar
}\right)  ^{N/4}\widehat{T}(-z_{0})e^{-\frac{1}{\hbar}|x|^{2}};\label{cohab}%
\end{equation}
where $\widehat{T}(z_{0})=e^{-\frac{i}{\hbar}(x_{0}\widehat{p}-p_{0}%
\widehat{x})}$ is the Heisenberg--Weyl operator. These states are minimum
uncertainty states (they saturate the Heisenberg inequalities $\Delta
x_{j}\Delta p_{j}\geq\frac{1}{2}\hbar$). A standard calculation of Gaussian
integrals shows that the scalar product of these states is
\begin{equation}
\langle\theta|\psi\rangle=e^{-\frac{1}{\hbar}|z_{0}|^{2}}\text{.}\label{kaxi}%
\end{equation}
Let us calculate $W(\phi,\psi)$. Using the translation formula (see de Gosson
\cite{Birkbis})
\begin{equation}
W(\widehat{T}(\alpha)\phi,\widehat{T}(\beta)\psi)(z)=e^{-\frac{i}{\hbar}%
\chi_{\alpha\beta}(z)}W(\phi,\psi)(z-\tfrac{1}{2}(\alpha+\beta))\label{wab}%
\end{equation}
where $\chi_{\alpha\beta}$ is the phase function defined by
\begin{equation}
\chi_{\alpha\beta}(z)=\tfrac{1}{2}\sigma(z,\alpha-\beta)+\sigma(\alpha
,\beta)\label{ki}%
\end{equation}
($\sigma$ the standard symplectic form). We thus have%
\[
W(\phi,\psi)(z)=e^{\frac{i}{\hbar}\sigma(z,z_{0})}W(\xi_{0},\xi_{0})(z)
\]
where $\sigma(z,z_{0})=px_{0}-p_{0}x$ and $\xi_{0}(x)=\left(  \pi\hbar\right)
^{-N/4}e^{-|x|^{2}/\hbar}$ is the standard fiducial coherent state (Littlejohn
\cite{Littlejohn}). Now, $W(\xi_{0},\xi_{0})=W\xi_{0}$, the Wigner
distribution of $\xi_{0}$, which is given by
\begin{equation}
W\xi_{0}(z)=\left(  \tfrac{1}{\pi\hbar}\right)  ^{N}e^{-\frac{1}{\hbar}%
|z|^{2}}\text{ \ , \ }|z|^{2}=|x|^{2}+|p|^{2}\label{wo}%
\end{equation}
(de Gosson \cite{Birk,Birkbis}, Littlejohn \cite{Littlejohn}). We thus
conclude that%
\begin{equation}
W(\phi,\psi)(z)=\left(  \tfrac{1}{\pi\hbar}\right)  ^{N}e^{\frac{i}{\hbar
}\sigma(z,z_{0})}e^{-\frac{1}{\hbar}|z|^{2}}\text{.}\label{walbe}%
\end{equation}
Using the scalar product formula (\ref{kaxi}) we see that the complex
probability distribution $\rho_{\phi,\psi}$ is given by%
\begin{equation}
\rho_{\phi,\psi}(z)=\left(  \tfrac{1}{\pi\hbar}\right)  ^{N}e^{\frac{i}{\hbar
}\sigma(z,z_{0})}e^{\frac{1}{\hbar}|z_{0}|^{2}}e^{-\frac{1}{\hbar}|z|^{2}%
}.\label{rab}%
\end{equation}
This formula shows that $\rho_{\alpha,\beta}(z)$ has an\ oscillatory behavior
which is sharply peaked near the origin. We notice that since%
\[
|\rho_{\phi,\psi}(z)|\leq\left(  \tfrac{1}{\pi\hbar}\right)  ^{N}e^{\frac
{1}{\hbar}|z_{0}|^{2}}e^{-\frac{1}{\hbar}|z|^{2}}%
\]
the weak value $\langle\widehat{A}\rangle_{\mathrm{weak}}^{\phi,\psi}$
satisfies
\begin{align*}
|\langle\widehat{A}\rangle_{\mathrm{weak}}^{\phi,\psi}| &  \leq\int%
_{\mathbb{R}^{2N}}|\rho_{\phi,\psi}(z)||A(z)|dz\\
&  =\left(  \tfrac{1}{\pi\hbar}\right)  ^{N}e^{\frac{1}{\hbar}|z_{0}|^{2}}%
\int_{\mathbb{R}^{2N}}e^{-\frac{1}{\hbar}|z|^{2}}|A(z)|dz\\
&  \leq\left(  \tfrac{1}{\pi\hbar}\right)  ^{N}e^{\frac{1}{\hbar}|z_{0}|^{2}%
}\sup|A(z)|\int_{\mathbb{R}^{2N}}e^{-\frac{1}{\hbar}|z|^{2}}dz.
\end{align*}
The integral in the third line is easy to evaluate; its value is $(\pi
\hbar)^{N}$ hence we have the estimate%
\begin{equation}
|\langle\widehat{A}\rangle_{\mathrm{weak}}^{\phi,\psi}|\leq e^{\frac{1}{\hbar
}|z_{0}|^{2}}\sup|A(z)|.\label{estima}%
\end{equation}
This inequality shows that even if the observable $A$ is small, the weak value
can a priory take very large values provided that the phase space distance
between both wavepackets $\phi,\psi$ is large; this is in strong contrast with
what happens for the individual states $|\phi\rangle$ and $|\psi\rangle$, for
which lead to the estimates%
\[
|\langle\widehat{A}\rangle^{\phi}|\leq\sup|A(z)|\text{ \ , \ }|\langle
\widehat{A}\rangle^{\psi}|\leq\sup|A(z)|;
\]
the relative phase space localization of these states does not play any role
in these inequalities. We will shortly discuss non-trivial extensions of the
superposition considered above in the discussion below.

\section{Discussion}

Let us apply the phase space formalism to a discussion of the situation
initially considered in \cite{avphysa,av} where at a time $t_{\mathrm{in}}$ an
observable $\widehat{A}$ is measured and a non-degenerate eigenvalue was
found: $|\psi(t_{\mathrm{in}})\rangle=|\widehat{A}=a\rangle$ (the pre-selected
state); similarly at a later time $t_{\mathrm{fin}}$ a measurement of another
observable $\widehat{B}$ yields $|\phi(t_{\mathrm{fin}})\rangle=|\widehat{B}%
=b\rangle$ (the post-selected state). Let $t$ be some intermediate time:
$t_{\mathrm{in}}<t<t_{\mathrm{fin}}$. Following the time-symmetric approach to
quantum mechanics (see the review in \cite{apt}), at this intermediate time
the system is described by the \emph{two} wavefunctions%
\begin{equation}
\psi_{t}=U_{t,t_{\mathrm{in}}}^{H}\psi(t_{\mathrm{in}})\text{ \ , \ }\phi
_{t}=U_{t,t_{\mathrm{fin}}}^{H}\phi(t_{\mathrm{fin}})\label{t12}%
\end{equation}
where $U_{t,t^{\prime}}^{H}=e^{-i\widehat{H}(t-t^{\prime})/\hbar\text{ }}$ is
the Schr\"{o}dinger unitary\ evolution operator ($\widehat{H}$ the quantum
Hamiltonian). Notice that $\phi_{t}$ travels \emph{backwards} in time since
$t<t_{\mathrm{fin}}$. The situation is thus the following: at any time
$t^{\prime}<t$ the system under consideration is in the state $|\psi
_{t^{\prime}}\rangle=U_{t^{\prime},t_{\mathrm{in}}}^{H}|\psi(t_{\mathrm{in}%
})\rangle$ and has Wigner distribution $W\psi_{t^{\prime}}$; at any time
$t^{\prime\prime}>t$ the system is in the state $|\phi_{t^{\prime\prime}%
}\rangle=U_{t^{\prime\prime},t_{\mathrm{fin}}}^{H}|\phi(t_{\mathrm{in}%
})\rangle$ and has Wigner distribution $W\phi_{t^{\prime}}$. But at time $t$
it is the superposition $|\psi_{t}\rangle+|\phi_{t}\rangle$ of both states,
and the Wigner distribution of this cat-like state is
\begin{equation}
W(\phi_{t}+\psi_{t})=W\phi_{t}+W\psi_{t}+2\operatorname{Re}W(\phi_{t},\psi
_{t}).\label{wpsifi}%
\end{equation}
This equality shows the abrupt emergence at time $t$ --and only at that
time!-- of the interference term $2\operatorname{Re}W(\phi_{t},\psi_{t})$,
signalling a strong interaction between the states $|\psi_{t}\rangle$ and
$|\phi_{t}\rangle$. Such an interaction is due to the wavelike nature of
quantum mechanics, and is absent from classical mechanics. The appearance of
interference terms described by the cross-Wigner transform is well-known and
considered as an asset in time-frequency analysis (e.g. radar theory, see
Cohen \cite{Cohen}, Auslander and Tolimieri \cite{aus}). It seems therefore
that our approach could well open new perspectives in the topic of weak
measurements and values, by importing robust techniques from these Sciences
(it is a fact, due mainly to historical and technical reasons, that the
mathematical techniques related to the Wigner formalism have grown faster and
are more sophisticated in signal theory and time-frequency analysis than they
are in quantum mechanics, so a feedback seems to be more than welcome!).

How the weak values are related to sub-Planckian scales would also be
interesting to investigate; the discussion in Zurek \cite{zurek}, and
especially the results in Nicacio et al. \cite{nic} could certainly be useful
in this context. These authors consider superpositions of an arbitrary number
of Gaussian states, and study their motion under the action of arbitrary
Hamiltonian flows. They show that the interference terms coming from the
cross-Wigner transforms are always hyperbolic and survive the action of a
thermal reservoir. While they mainly have in mind semiclassical dynamics,
their approach could be implemented in the context of weak values. It is
actually to a large extent sufficient to study the case of coherent states as
in Subsection \ref{subco}, because these states form an overcomplete set in
the square-integrable functions. In fact, choosing an adequate lattice
$\Lambda$ of points $z_{0}$ in phase space the functions $\widehat{T}%
(z_{0})\xi_{0}$ ($\xi_{0}(x)=\left(  \pi\hbar\right)  ^{-N/4}e^{-|x|^{2}%
/\hbar}$) form a Gabor frame (Gr\"{o}chenig \cite{Gro}) allowing to write an
arbitrary pure state as a linear superposition of the states $\widehat{T}%
(z_{0})\xi_{0}$. The net contribution of all cross-Wigner transforms of pairs
$(\widehat{T}(z_{0})\xi_{0},\widehat{T}(z_{1})\xi_{0})$ with $z_{0}\neq z_{1}$
is then the total interference leading to weak values (in \cite{zurek} Zurek
considers a \textquotedblleft compass state\textquotedblright\ consisting of
four terms $\widehat{T}(z_{0})\xi_{0}$, of which he studies interference
effects at the sub-Planckian scale; it would be interesting to interpret his
results in terms of weak values).

There is another aspect of the theory of weak values we have not mentioned at
all, if only because of lack of space and time. It is the possibility of
reconstructing wave functions from weak values, as initiated in Lundeen et al.
\cite{lund}. It turns out that the Wigner approach sketched in this Letter
leads to useful formulas. For instance, on proves the following inversion
formula (de Gosson \cite{Birkbis}, \S 9.4.2): Let $\eta$ be an arbitrary
square integrable function such that $\langle\phi|\gamma\rangle\neq0$; then%
\begin{equation}
\psi(x)=\frac{2^{N}}{\langle\phi|\gamma\rangle}\int_{\mathbb{R}^{2N}}%
W(\phi,\psi)\langle\widehat{T}_{\text{GR}}(z_{0})\psi|\gamma\rangle
dz_{0}.\label{inv1}%
\end{equation}
We can reconstruct $\psi$ from the knowledge of the weak value provided that
we know $\langle\phi|\gamma\rangle$. This inversion formula together with the
notion of mutually unbiased bases (MUB) could certainly play an important role
in the reconstruction problem.\bigskip

\noindent\textbf{Acknowledgements}. The first author (MdG) has been supported
by the EU FET Open grant UNLocX (255931). Both authors wish to thank Basil
Hiley (Birkbeck) for useful remarks and helpful criticism. We are also happy
to thank Hans Feichtinger for having pointed out several misprints.

\end{document}